\documentstyle[epsfig]{aipproc2}
\def\vev#1{\left\langle #1\right\rangle}

\begin{document}
\title{
\begin{flushright} \small
FTUV/99-13 \\
IFIC/99-13 \\
UWThPh-1999-12\\
FSU-HEP-030399
\end{flushright}
R-parity violating decays of the Top-Quark and the Top-Squark
       at the Tevatron\footnote{\scriptsize To appear in the  
proceedings of the workshop ``Physics at Run II: Supersymmetry/Higgs Fermilab, 
Feb.--Nov. 1998''}}
\author{F.~de Campos$^1$,
        M.~A.~D\'\i az$^2$,
        O.~J.~P.~Eboli$^4,6$,
	M. B. Magro$^6$,
        L.~Navarro$^3$,
        W.~Porod$^5$,
        D.~A.~Restrepo$^3$, and
        J.~W.~F.~Valle$^3$
           }
\address{ 
       $^1$Departamento de F\'\i sica y Qu\'\i mica, Univ. Estadual Paulista, 
       Guaratinguet\'a, Brasil \\
       $^2$ Department of Physics, Florida State University, Tallahassee,
        FL 32306, USA\\
       $^3$Departamento de F\'\i sica Te\'orica, IFIC-CSIC, Univ.~de Valencia,
            Spain \\ 
     $^4$Instituto de F\'\i sica Te\'orica, Univ. Estadual Paulista,
          Sao Paulo, Brasil \\
       $^5$Insitut f\"ur Theoretische Physik, Univ.~Wien, Austria\\
$^6$ Physics Department, University of Wisconsin, 1150 University Av. 
Madison, WI 53706, USA
        }
\maketitle
\abstract{We study unconventional decays of the top-quark and the top-squark
in the framework of SUSY models with broken R-parity. The model under
study is the MSSM with an additional bilinear term that breaks
R-parity. In this model the top-squark behaves similar to a third
generation leptoquark.  We demonstrate that existing Tevatron data on
the top give rise to restrictions on the SUSY parameter space. In
particular, we focus on scenarios where the tau-neutrino mass is
smaller than 1 eV. We give an exclusion plot derived from the
leptoquark searches at Tevatron.}

\section{Introduction}

The search for supersymmetry (SUSY) is one of the main tasks in the
experimental program of the Tevatron. Most of the studies have been carried
out in the framework of the Minimal Supersymmetric Standard Model (MSSM)
(see e.g. \cite{carena} and references therein). There has also
been considerable work in the case of R-parity violation \cite{old}
(for collider studies see e.g.~\cite{frenchreport} and references therein). 
 The latter ones have mainly treated the case of trilinear R-parity
violating couplings (TRPV).
Here we focus on the case of bilinear R--Parity Violation
(BRPV)~\cite{epsrad,BRPVtalk,BRPhiggs,others} which contain as additional
feature a vev for the sneutrinos.  These models are
well-motivated theoretically as they arise as effective truncations of
models where R--Parity is broken spontaneously \cite{SRpSB} through
right handed sneutrino vacuum expectation values (vev)
$\vev{\tilde\nu^c}=v_R \neq0$. 
They open new possibilities for the study of the unification
of the Yukawa couplings~\cite{epsbtaunif}. In particular it has been
shown that in BRPV models bottom-tau unification may be
achieved at any value of $\tan\beta$. 
From a phenomenological point of view these models predict a plethora
of novel processes \cite{BRPVtalkphen} that could reveal the existence
of SUSY in a totally different way, not only through the usual missing
momentum signature as predicted by the MSSM.
They provide a very predictive approach to the
violation of R--Parity, which renders the systematic study of
R-parity violating physics \cite{BRPVtalkphen} possible.
Moreover, they are more
restrictive than TRPV models, especially in their supergravity
formulation, if universality of the soft-breaking terms is assumed
at the unification scale, as in ~\cite{epsrad}.

We will consider the simplest superpotential which violates R-Parity
\begin{equation} 
W_{R_p \hspace{-3.4mm} /} = W_{MSSM} 
+\epsilon_i \widehat L_i \widehat H_u \,,
\label{eq:Wsuppota}
\end{equation}
assuming that TRPV terms are absent or suppressed, as would be
the case if their origin is gravitational~\cite{BJV}.  The
$\epsilon_i$ terms violate lepton number in the $i$th generation
respectively. As already mentioned, models where R--Parity is broken
spontaneously \cite{SRpSB} through a vev of the
right handed sneutrinos $\vev{\tilde\nu^c}=v_R \neq0$ generate only
BRPV terms. The $\epsilon_i$ parameters are then identified as a product of
a Yukawa coupling and $v_R$. This provides the main
theoretical motivation for introducing explicitly BRPV in the MSSM
superpotential.
For simplicity we set from now on $\epsilon_1=\epsilon_2=0$, and in this
way, only tau--lepton number is violated. In this case, considering
only the third generation, the MSSM--BRPV has the following
superpotential
\begin{equation} 
W_{R_p \hspace{-3.4mm} /}=\varepsilon_{ab}\left[
 h_t\widehat Q_3^a\widehat U_3\widehat H_u^b
+h_b\widehat Q_3^b\widehat D_3\widehat H_d^a
+h_{\tau}\widehat L_3^b\widehat R_3\widehat H_d^a
+\mu\widehat H_u^a \widehat H_d^b
+\epsilon_3\widehat L_3^a\widehat H_u^b\right]\,,
\label{eq:Wsuppot}
\end{equation}
where the first four terms correspond to the MSSM. The last term violates
tau--lepton number as well as R--Parity. 

\noindent
\begin{minipage}[t]{65mm}
{\setlength{\unitlength}{1mm}
\begin{picture}(72,77)
\put(-3,2){\mbox{\psfig{figure=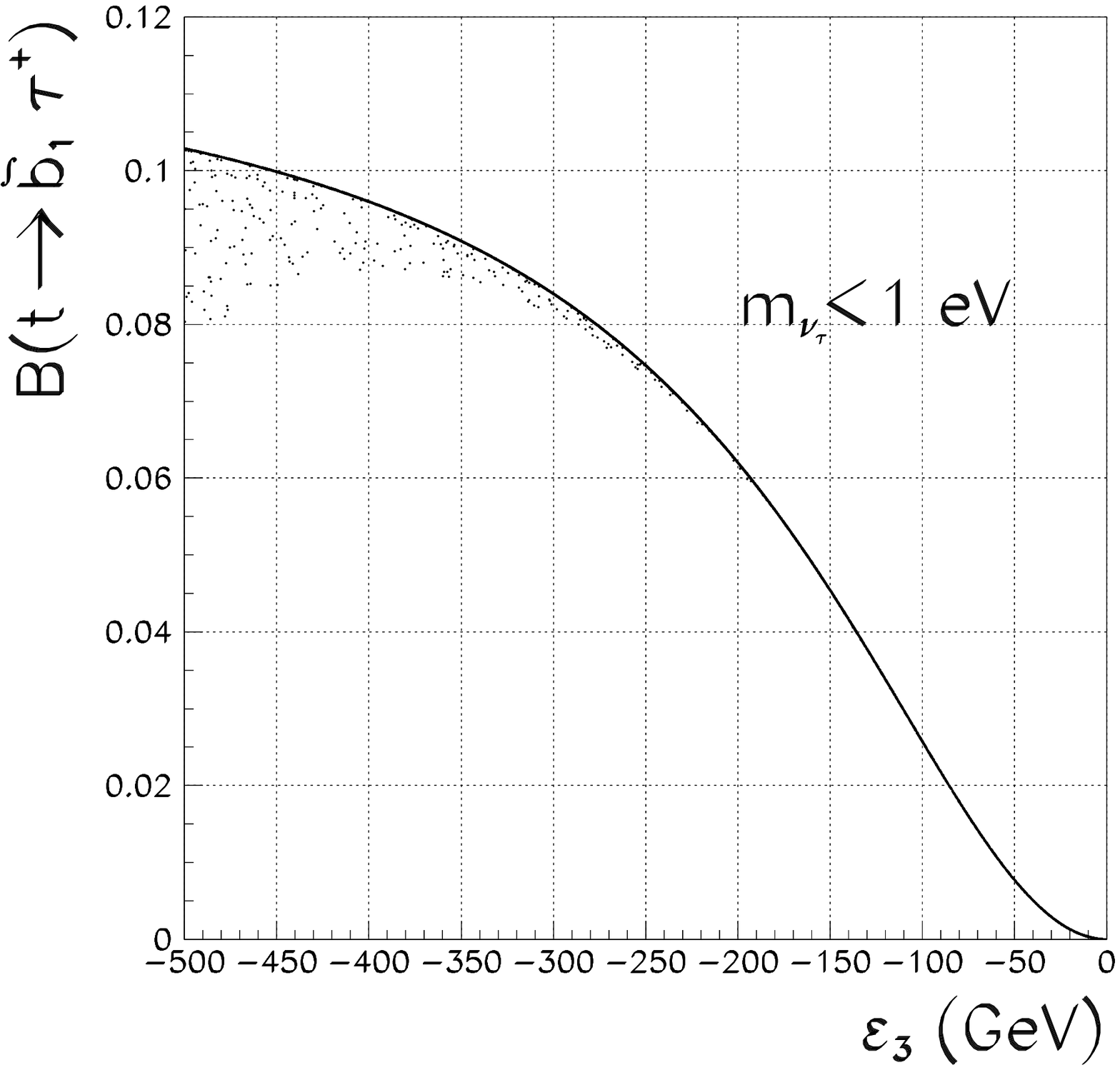,
                        height=7.3cm}
}}
\end{picture}}
\refstepcounter{figure}
\label{TopSbotTau}
{\small{\bf{Fig.~\ref{TopSbotTau}:}}
Branching Ratios for $t \to {\tilde b}_1 \, \tau^+$ as a
         function of  $\epsilon_3$. The parameters are:
         $M = 180$~GeV, $\mu = 200$~GeV, $\tan \beta = 35$,
         $M_{E_3} = 285$~GeV, $A_\tau = 280$~GeV, $M_Q = 285$~GeV,
         $M_U = 180$~GeV, $M_D = 190$~GeV, $A_t =320$~GeV,
         $A_b = 120$~GeV, $B = 50$~GeV, $-500$~GeV $< \epsilon < 0$~GeV,
         $0$~GeV $< B_2 < 200$~GeV,  $1$~GeV $< v_3 < 50$~GeV.}
\end{minipage}
\hspace{2mm}
\noindent
\begin{minipage}[t]{65mm}
{\setlength{\unitlength}{1mm}
\begin{picture}(72,77)
\put(-2,1){\mbox{\psfig{figure=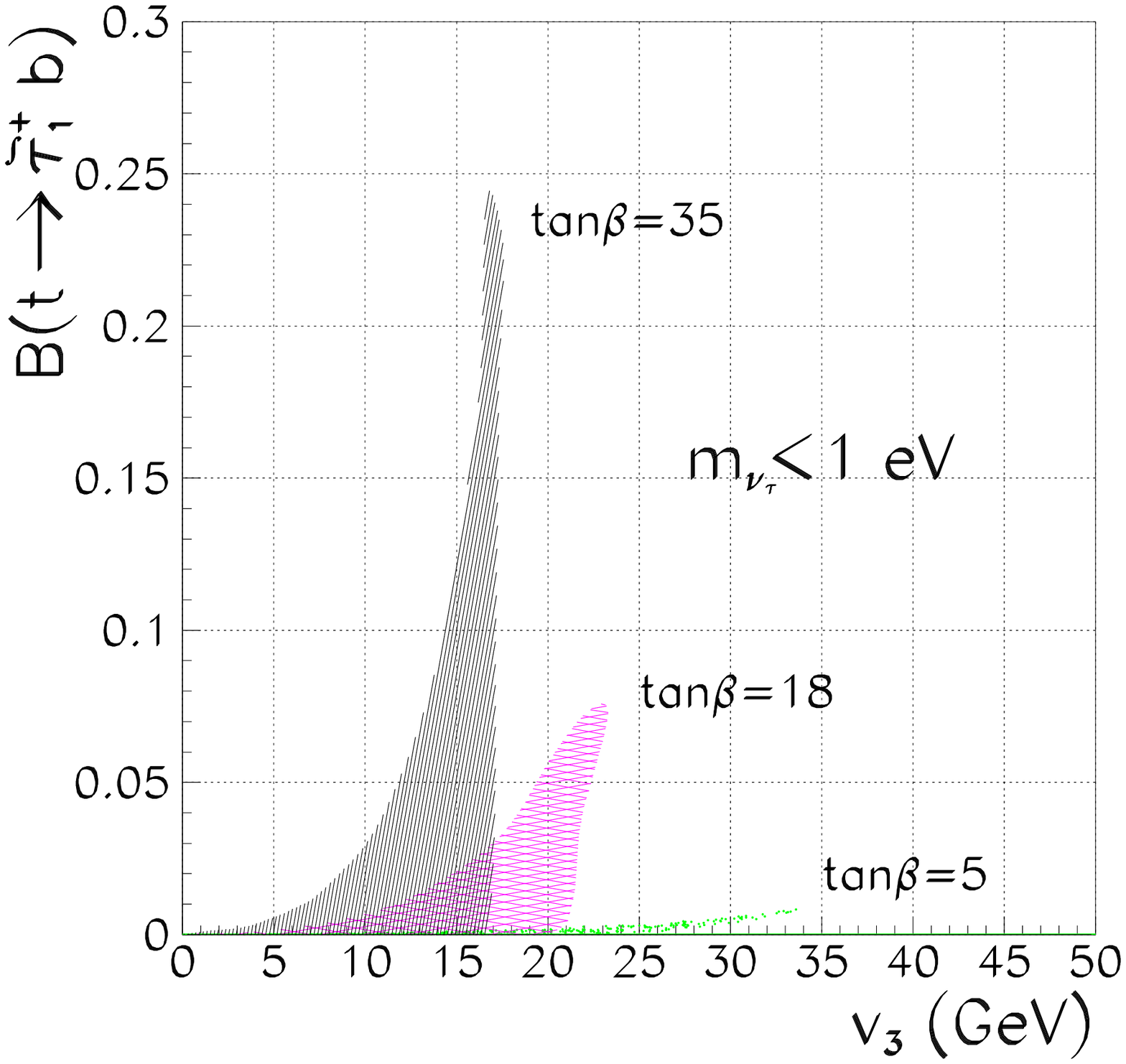,height=7.3cm}}}
\end{picture}}
\refstepcounter{figure}
\label{TopStauB}
{\small{\bf{Fig.~\ref{TopStauB}:}}
Branching Ratios for $t \to {\tilde \tau}^+_1 \, b$ as a
         function of  $v_3$ for different values of $\tan \beta$. 
         The other parameters are the same as in Fig.~\ref{TopSbotTau}.}
\end{minipage}
\vspace{4mm}

It has often been claimed that the BRPV term can be rotated away from
the superpotential by a suitable choice of the basis
\cite{HallSuzuki}.  If this were true then the $\epsilon$ term would
be unphysical.  Indeed, one can show that, performing this rotation of
the superfields one can indeed eliminate the BRPV but RPV is
reintroduced in the form of TRPV.  Moreover, supersymmetry must be
broken and the presence of the $\epsilon$ term in the superpotential
also introduces R-parity a violating term $\varepsilon_{ab} (B_2
\epsilon_3 L_3^a H^b_u)$, in the scalar sector implying that the
vacuum expectation value $\vev{\tilde\nu_{\tau}}=v_3/\sqrt{2}$ is
non--zero. This in turn generates more R--parity and tau lepton number
violating terms inducing a tau neutrino mass. Moreover, it is in
general impossible to rotate away the bilinear term in the
Superpotential and at the soft SUSY breaking potential at the same
time.

In this model the top-quark as well as the top-squark get additional
decay modes, e.g. $t \to {\tilde \tau}^+_1 \, b$ or ${\tilde t}_1 \to
\tau^+ + b$.  We study these decay in view of the Tevatron (top decays
in TRPV models has been treated in \cite{dreiner}). We show that
existing Tevatron data give additional constraints on the parameter
space.

\section{Top Decays}

One of the major successes of Tevatron has been the discovery of the
top-quark \cite{Topdiscovery}. The large top mass implies a relatively
small production cross section at the Tevatron. Therefore, the sum of
all branching ratios of the top decays except $t \to W^+ \, b$ is only
restricted to be smaller than approximately 25 \% \cite{TopEx}.  In
the MSSM the top can decay according to: $t \to W^+ \, b$, $t \to H^+
\, b$, $t \to {\tilde \chi}^0_1 \, {\tilde t}_1 $, $t \to {\tilde
\chi}^+_1 \, {\tilde b}_1 $.
The last mode is only listed for completeness, because it is
practically ruled out by existing LEP2--data \cite{LepTalk}. In the
BRPV model the charginos mix with the charged leptons, the neutralinos
with neutrinos, and the charged sleptons with the charged Higgs boson
\cite{epsrad,BRPVtalk,BRPhiggs}. Therefore, the top can have
additional decay modes:
\begin{equation}
t \to {\tilde \tau}^+_1 \, b \, , \hspace{4mm}
t \to \nu_\tau \, {\tilde t}_1 \, , \hspace{4mm}
t \to \tau^+ \, {\tilde b}_1 \, .
\end{equation}
As an illustrative example we show in Fig.~\ref{TopSbotTau} the
branching ratio for $t \to \tau^+ \, {\tilde b}_1 $ as a function of
 $\epsilon_3$. We have randomly chosen 10000 points
imposing the following constraints: $m_{\nu_\tau} < 18$~MeV, $m_{
{\tilde t}_1}, m_{ {\tilde b}_1} > 80$~GeV, $min(m_{H^+}, m_{ {\tilde
\tau}_1}) > 70$~GeV, and $m_{{\tilde \chi}^+_1} > 85$~GeV. The
parameters are listed in the figure caption. We find a strong
correlation between the R-parity decay branching ratio $BR(t \to
\tau^+ \, {\tilde b}_1)$ and the magnitude of $\epsilon_3$. 
This can be understood in the following way: in the chargino mass
matrix the mixing between the leptons and the charginos disappears if
one does the following rotation in the superfields: $\widehat H_1 \to
N (\mu \widehat H_1 - \epsilon_3 \widehat L_3)$ and $\widehat L_3 \to
N (\mu \widehat L_3 + \epsilon_3 \widehat H_1)$ (N being the
normalization). In this basis the coupling between $t$, $\tau$, and
${\tilde b}_1$ is proportional $N \, h_b \,\epsilon_3$ leading to this
feature.  

In Fig.~\ref{TopStauB} we show the branching ratio for $t \to {\tilde
\tau}^+_1 \, b$ as a function of $v_3$. Results in Fig.~\ref{TopStauB} are displayed for different values of
$\tan \beta$ and the other parameters are also the same as in Fig.~\ref{TopSbotTau} . The
dependence on $\tan \beta$ is a result of: (i) The stau - charged
Higgs boson mixing is proportional to the R-parity breaking parameters
$\epsilon_3$ and $v_3$ (ii) The decay width depends on the bottom
Yukawa coupling which increases with $\tan \beta$.  As can be seen
from the figure there is a strong correlation between the magnitude of
the R-parity breaking branching ratios and the mixing between the stau
and the charged Higgs boson.

We have performed a similar scan for small $\tan \beta$ for both of
BRPV decay channels discussed above.  These are suppressed in this case
and can not exceed 2\% or so, i.e. ($\sum BR(t \to b \, X) < 1-2\% \,
(X\ne W)$), because their decay widths are in all cases proportional
to the bottom Yukawa coupling squared.  In the case of $t \to \tau^+
\, {\tilde b}_1$ this is clear from the discussion of
Fig.~\ref{TopSbotTau}.  In the case of $t \to {\tilde \tau}^+_1 \, b$
one has to note that the stau mixes mainly with charged component of
the down-type Higgs multiplet $H_1$ ($\tilde \tau_L$ and $H_1$ have
the same gauge quantum numbers) and the $H_1 t b$ coupling is
proportional to $h_b$.
 
In every case the various decay modes lead to cascade decays:\\
\\
\begin{tabular}{lll}
$t \to {\tilde \tau}^+_1 \, b$ & $\to \tau^+ \, \nu_\tau \, b$ & \\
           & $\to \tau^+ \, {\tilde \chi}^0_1 \, b$ &
                      $\to \tau^+ \, f \, \bar{f} \, \nu_\tau \, b$ \\
           & &   $\to \tau^+ \, f \, \bar{f}' \, \tau^\pm \, b$ \\
           & $\to \nu_\tau \, {\tilde \chi}^+_1 \, b$ &
                      $\to \nu_\tau \, f \, \bar{f}' \, \nu_\tau \, b$ \\
           & &   $\to \nu_\tau \, f \, \bar{f} \, \tau^+ \, b$ \\
           &  $\to c \, s \, b$ & \\
$t \to \tau^+ \, {\tilde b}_1$ & $\to \tau^+ \, \nu_\tau \, b$ & \\
           & $\to \tau^+ \, {\tilde \chi}^0_1 \, b$ &
                      $\to \tau^+ \, f \, \bar{f} \, \nu_\tau \, b$ \\
           & &   $\to \tau^+ \, f \, \bar{f}' \, \tau^\pm \, b$ 
\end{tabular} 
\vskip .2cm
In nearly all cases there are two $\tau$'s and two $b$-quarks in the
final state plus the possibility of additional leptons and/or jets.
Therefore, $b$-tagging and a good $\tau$ identification are important
for extracting these final states. Moreover there is in general a
large multiplicity of charged particles in the final state which
should be helpful in reducing the background. The background will come
mainly from the production of one or two gauge bosons plus
additional jets.
The conclusion in similar cases \cite{LeCompte98} has been that in its
next run the Tevatron should be sensitive to branching ratio values up
to $10^{-3} - 10^{-2}$ depending on the mode. Therefore, the possible
observation of one of these additional decay modes at the run 2 of
Tevatron should give a strong hint on the underlying parameters.

\section{Top-Squark Decays}

Top-squark physics is a very interesting part of supersymmetric
theories, because the lighter top-squark might be the lightest charged
SUSY particle.  This follows because: (i) The large top Yukawa
coupling leads to reduced soft SUSY breaking masses compared to the
first two generation in GUT models (see e.g. \cite{Drees} and
references therein), and (ii) The off-diagonal element of the
top-squark mass matrix is proportional to the top mass leading to a
strong mixing and possible light mass eigenstate.

In the kinematical reagion accessible to the Tevatron the light
top-squark has the following MSSM decay modes: $ {\tilde t}_1 \to
{\tilde \chi}^+_i + b$, $ {\tilde t}_1 \to {\tilde \chi}^0_1 + c$, $
{\tilde t}_1 \to {\tilde l}^+_i + \nu_l + b$, $ {\tilde t}_1 \to
{\tilde \nu}_l + l^+ + b$, $ {\tilde t}_1 \to {\tilde \chi}^0_1 + W^+
+ b$, and $ {\tilde t}_1 \to {\tilde \chi}^0_1 + H^+ + b$ (for a
discussion see e.g. \cite{StopDecays} and references therein).  In
BRPV models the top-squark has an additional and phenomenologically
very interesting decay mode \cite{stop}:
\begin{equation}
 {\tilde t}_1 \to \tau^+ + b
\end{equation}
In the following we have concentrated on scenarios where only the
two-body decay modes are possible. We adopt the framework of
Supergravity unification \cite{epsrad} in order to reduce the number
of free SUSY parameters. However, we keep $\epsilon_3$ and $v_3$ as
free parameters for the moment. In Fig.~\ref{figstop} we show the
areas in the $m_{{\tilde t}_1}$-$m_{{\tilde \chi}^0_1}$ plane where
the branching ratio ${\tilde t}_1 \to \tau^+ + b$ is larger than 90\%
for different values of $\epsilon_3$ and $v_3$. We restrict to the
range $|\epsilon_3|, |v_3|<1$ GeV, and vary randomly the MSSM
parameters keeping $m_{\tilde t_1}<m_{{\tilde \chi}^{\pm}_1}+m_b$.
This demonstrates that one can get a dominance of the R-Parity
violating decay mode even for relatively small values of the R-parity
breaking parameters.  The upper--left triangular region corresponds to
$m_{\tilde t_1}<m_{{\tilde \chi}^0_1}+m_c$ and thus $BR(\tilde t_1 \to
b \tau)=1$. In the lower--right triangular region $m_{\tilde
t_1}>m_{{\tilde \chi}^+_1}+m_b$ and therefore $\tilde t_1 \to b
\, {\tilde \chi}^+_1$ is open. In the central region the top-squark
has the two decay modes $\tilde t_1 \to b \, \tau$ and $\tilde t_1 \to
c \, {\tilde \chi}^0_1$. The solid lines, defined by the maximum value
of $|\epsilon_3|$ and $|v_3|$, are the boundary of the regions where
$BR(\tilde t_1 \to b \tau)>0.9$ such that points at the left of the
boundary satisfy that condition.

Since BRPV models allow the decay $({\tilde t}_1 \to \tau^+ + b)$ we
can interpret the top squark as a third generation leptoquark.
Therefore we can use the limits obtained from leptoquark searches
\cite{LQ} to derive limits on the top-squark for this case. In
Fig.~\ref{stopexclude} we show an exclusion plot in the
$m_0$-$m_{1/2}$ plane. The nearly horizontal dashed lines are chargino
mass contours and the lines forming radial patterns are the top-squark
mass contours.  The upper to the lower radial curves corresponds to
$m_{\tilde{t}_1}=120, 100$ and 80. The region limited by the
dotted-dashed line is defined by $m_{\tilde{t}_1}\;<\;m_{{\tilde
\chi}^+_1}$.  The analysis rules out $m_0$ and $m_{1/2}$ points in the
dark hashed region. In the lower hashed region no points with
radiative electroweak symmetry breaking can be found. We have taken
$\tan \beta= 3$, $A_0=-650$ GeV and $\epsilon_3 / \mu = -0.5$ and
verified that in this region $BR(\tilde{t}_1 \to b \tau)=1$.  The
Tevatron limits can not be directly applied when $A_0>-500$ GeV,
because in this case $m_{\tilde{t}_1}\;>\;m_{{\tilde \chi}^+_1}+m_b$.
The regions in the $m_0$--$m_{1/2}$ plane where 
$m_{\tilde{t}_1}\;<\;m_{{\tilde \chi}^+}$ are excluded if $-650 < A_0
< -500$ GeV and $|\epsilon_3 / \mu|$ is sufficiently large so that the
three-body decays are negligible.
Fig.~\ref{stopexclude} shows that the region where
$m_{\tilde{t}_1}\;<\;m_{{\tilde \chi}^+_1}+m_b$ and $BR({\tilde t}_1
\to \tau^+ + b) \approx 1$ is practically ruled out by experiment.
For this particular choice  of SUSY parameters there is only
a little window still to explore at the run 2 of Tevatron. However for
other choices of SUSY parameters, e.g. $A_0=-900$ GeV the dark-hatched
region fills up only about half of the allowed region where
$m_{\tilde{t}_1}\;<\;m_{{\tilde \chi}^+_1}+m_b$ and would therefore
be open for investigation at the next run.

The MSSM three-body channels could be competitive with the BRPV one if
$|\epsilon_3 / \mu|$ is very small and $m_0 \ll m_{1/2}$. In this case
the condition $BR({\tilde t}_1 \to \tau^+ + b) \approx 1$ no longer
holds and our analysis is not applicable.  If $|\epsilon_3 /
\mu|<10^{-3}$ (which leads to tau neutrino mass in the 10$^{-2}$eV
range in the mSUGRA model) for the same value of $\tan\beta$ we find 
\cite{staumad} that the decay mode into $c\, {\tilde \chi}^0_1$ is 
competitive with the $b \tau$ channel.
The $\tilde{t}_1 \to c  {\tilde \chi}^0_1$ channel becomes more
important for large $\tan \beta$ and $m_{\tilde{t}_1}\;<\;m_{{\tilde
\chi}^+_1}$. In this case one needs $|\epsilon_3 / \mu|>10^{-2}$ in 
order to get a negligible $BR(\tilde{t}_1 \to c {\tilde \chi}^0_1 )$.

\section{Summary}

We have studied top-quark and top-squark decays in a supersymmetric
model with bilinear R-parity breaking. We have found that in both
cases there exist additional top and stop decay modes leading to novel
phenomenological implications with respect to those of the MSSM. In
the top-quark case the new decay modes are $t \to {\tilde
\tau}^+_1 \, b$, $t \to \nu_\tau \, {\tilde t}_1$, and $t \to \tau^+
{\tilde b}_1$.  We have shown that existing data on non-W top decay
from Tevatron are already sensitive to the BRPV parameters, adding
both sbottom and stau decay channels. 

In this model the top-squark has the additional channel ${\tilde t}_1
\to \tau^+ + b$. This channel will be 100\% if the stop is the lightest 
SUSY particle, which is possible in the BRPV model. Moreover, we have
demonstrated that this decay can be dominant even when the lightest
neutralino below the stop and the R-parity breaking parameters
$|\epsilon_3|$ and $|v_3|$ are well below a GeV, as long as the
R-parity conserving chargino decay mode is kinematically closed,
i.e. for $m_{\tilde{t}_1}\;<\;m_{{\tilde \chi}^+_1}+m_b$. We have studied
scenarios in a SUGRA model with universality of the soft breaking
terms at the unification scale and we have found that the Tevatron
data on third generation lepto-quark data rule out scalar-top masses
below 80-100~GeV, depending on the parameters. Additional analysis
to determine the sensitivity region for this is now underway
\cite{stopanalysis}.

\section*{Acknowledgements}

This work was supported by DGICYT under grant PB95-1077 and Acci\'on
Integrada Hispano-Austriaca HU1997-0046, by the TMR network grant
ERBFMRX-CT96-0090 of the European Union, by CNPq and FAPESP (Brazil),
by Programa de Apoio a N\'ucleos de Excel\^encia (PRONEX), by a
CSIC-CNPq exchange agreement. D.~A.~R.~was supported by Colombian
COLCIENCIAS fellowship.  W.~P.~was supported by the ''Fonds zu
F\"orderung der wissenschaftlichen Forschung'' of Austria, project
No. P13139-PHY. L.~N. was supported by Spanish CSIC fellowship.

\newpage

\noindent
\begin{minipage}[t]{65mm}
{\setlength{\unitlength}{1mm}
\begin{picture}(72,77)
\put(-3,-1){\mbox{\psfig{figure=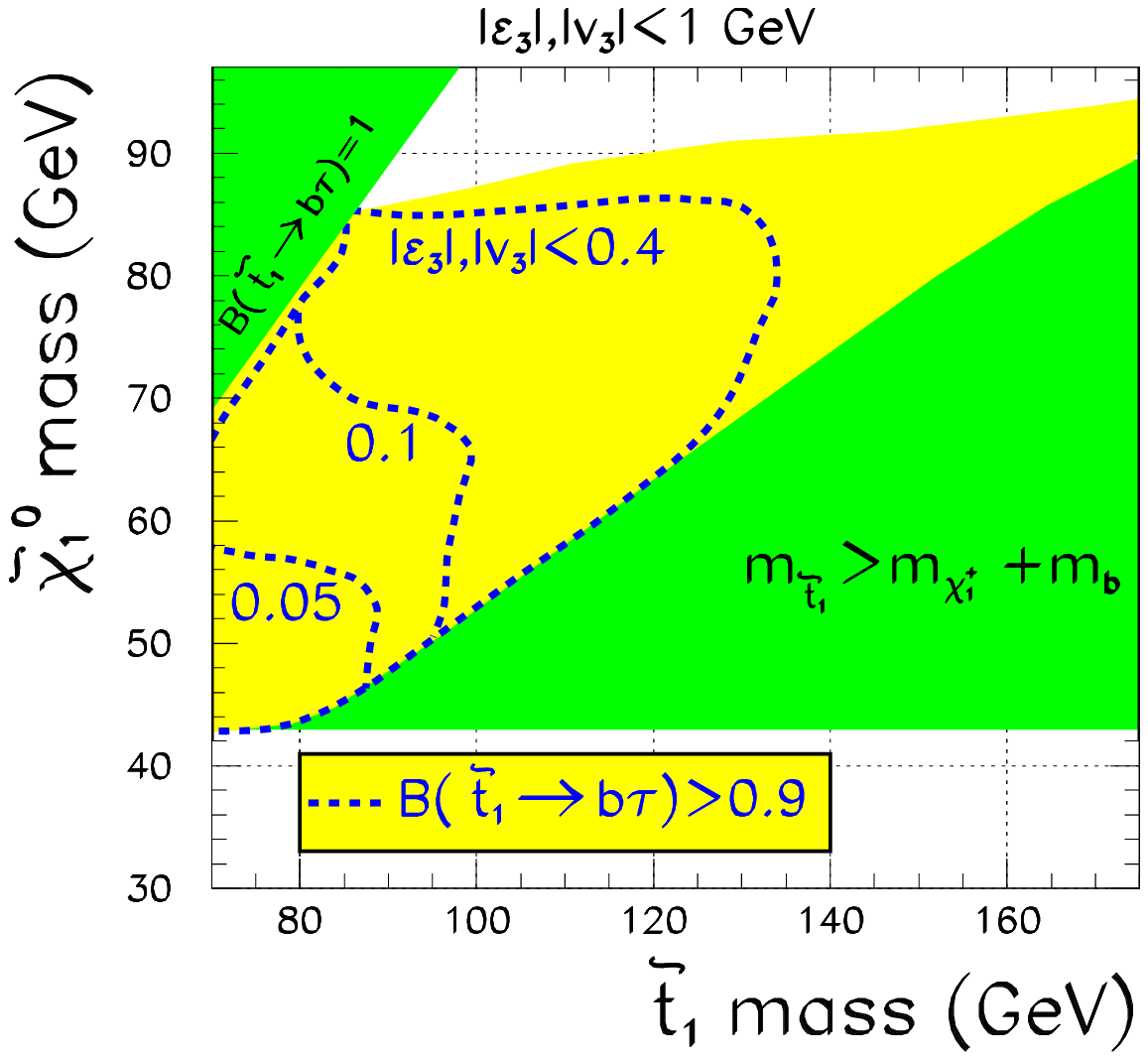,
                        height=7.0cm,width=7.5cm}}}
\end{picture}}
\refstepcounter{figure}
\label{figstop}
{\small{\bf{Fig.~\ref{figstop}:}} Contour-lines for BR$(\tilde t_1 \to
b \tau)>0.9$ in the $m_{{\tilde t}_1}$--$m_{{\tilde\chi}_1^0}$
plane. The gray region shows the area where only those two decay modes
are open. We consider $|\epsilon_3|, |v_3|<1$ GeV, and the MSSM
parameters are varied randomly such that $m_{\tilde t_1}<m_{{\tilde
\chi}^{\pm}_1}+m_b$.  The lines are defined by the maximum value of
$|\epsilon_3|$ and $|v_3|$ and delimit the regions where $BR(\tilde
t_1 \to b \tau)>0.9$.}
\end{minipage}
\hspace{3mm}
\noindent
\begin{minipage}[t]{65mm}
{\setlength{\unitlength}{1mm}
\begin{picture}(72,77)
\put(-3,-1){\mbox{\psfig{figure=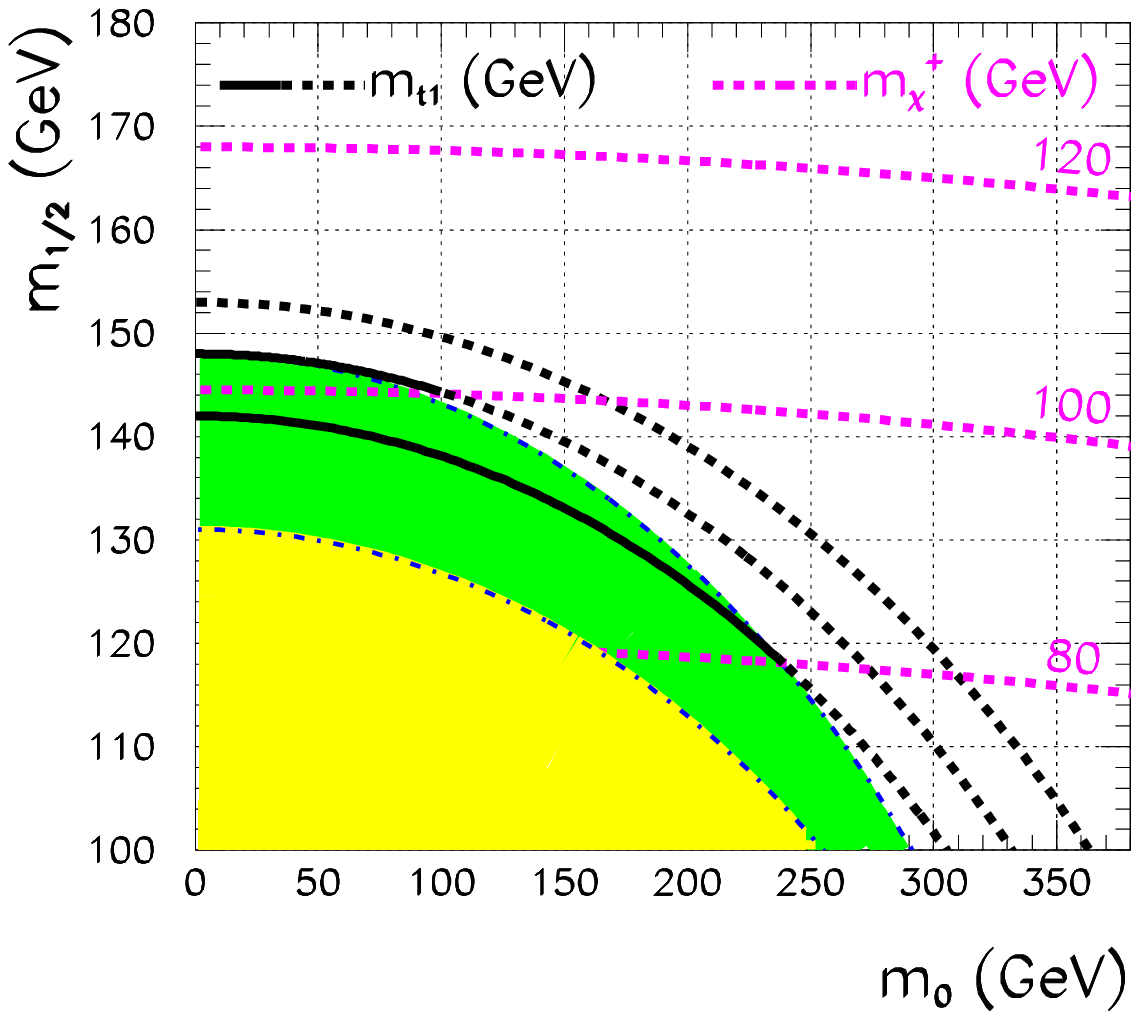,
                        height=7.0cm,width=7.5cm}}}
\end{picture}}
\refstepcounter{figure}
\label{stopexclude}
{\small{\bf{Fig.~\ref{stopexclude}:}} Exclusion contour in the
$m_0$--$m_{1/2}$ plane. The nearly horizontal dashed lines are
chargino mass contours while the radial-like dashed lines are the
top-squark mass contours. These change from solid to dashed when the
top--squark becomes heavier than the lightest chargino.  The radial
curves correspond to $m_{\tilde{t}_1}=120, 100$ and 80, respectively,
from upper to the lower. The region limited by the dotted-dashed line
has $m_{\tilde{t}_1}\;<\;m_{{\tilde \chi}^+}$.  The dark hashed region
is excluded by experimental data while the lower light-hashed region
is disfavoured by theory. We have fixed $\tan \beta= 3$, $A_0=-650$
GeV and $\epsilon_3 / \mu = -0.5$.}
\end{minipage}

\end{document}